# The Influence of Forebody Topology on Aerodynamic Drag and Aeroacoustics Characteristics of Squareback Vehicles using CAA

H. Viswanathan[*,**] and K.K Chode[*]
Corresponding author: h.viswanathan@shu.ac.uk

[*] Materials and Engineering Research Institute, Sheffield Hallam University, Howard Street, Sheffield, England, S1 1WB, United Kingdom.
[**] Department of Engineering and Mathematics, Sheffield Hallam University, Howard Street, Sheffield, England, S1 1WB, United Kingdom.

**Abstract:** This study numerically investigates the aerodynamic forces and flow-induced noise generated by SAE-T4, Ahmed, and Hybrid forebody shapes with a squareback vehicle configuration using SBES-FW-H. The results show significant differences in lift coefficients and the presence of a horseshoe vortex at the mirror, with smaller eddies that interact with A-pillar vortices, resulting in pronounced pressure fluctuations and noise generation on the side window for the three configurations. Surprisingly, negligible differences in aerodynamic drag and radiated sound are predicted despite these effects.

## 1  Introduction

The automotive industry continues to transform the aerodynamic and acoustics environment to alleviate discomfort, enhance communication systems, reduce vehicular emission noise such as pass-by noise signature, influence sound barriers on urban roads, and improve overall safety [1,2]. Vehicle noise sources can primarily be classified into three categories: aerodynamic noise, mechanical noise, and tire-road noise, with additional secondary classifications, such as slosh noise, occasionally gaining prominence [3-5]. In the current era of vehicle electrification, aerodynamic noise holds primary significance during cruising as wind-induced noise increases with the vehicle's speed and supersedes tire noise at around 100 km/h. Therefore, predicting and mitigating aerodynamic noise is critical to improving the driving experience and road safety.

Despite its potential significance, the impact of forebody on drag and its implications for sound generation and radiation from vehicles have been largely overlooked. To address this gap in the literature, this study investigates the impact of various forebody shapes on overall vehicle drag and its influence on the sound generated and radiated using computational aeroacoustics (CAA) method. The findings of this study expands the current knowledgebase on the relationship between the overall drag of different forebody shapes and aeroacoustics' performance of generic squareback vehicles.

## 2  System Description

In this study, three squareback vehicles with different forebody configurations were investigated, including the SAE-Type 4, Ahmed Body, and a Hybrid Body that combines features of the former two, as depicted in Fig. 1. To ensure comparability among the vehicle configurations studied, their height (h), length (L), reference areas (A), and with a generic mirror representing a square cylinder of length (d), and side window positions were kept identical. Each model was subjected to a freestream velocity of $U_\infty$ = 27.78 m/s, corresponding to a Reynolds number of $Re_L = 7 \times 10^6$ based on the length of the body. The simulation was setup to closely replicate the experimental conditions for the SAE-Type 4 configuration described in Refs. [6,7]. In each case, the models were placed in a computational domain of 12 L x 3.6 L x 3.6 L based on ERCOFTAC guidelines adapted from previous studies [8,9]. Additionally, the positions of the mirror and side window were matched to mitigate any extraneous variables that could confound the aerodynamic assessment and resulting sound generation and radiation characteristics of each configuration. The near-field flow was computed using the Stress Blended Eddy Simulation (SBES), which offers advantages over DDES and IDDES approaches, including stronger RANS layer shielding and a rapid, more definitive transition from RANS to LES models, even with unstructured meshes.

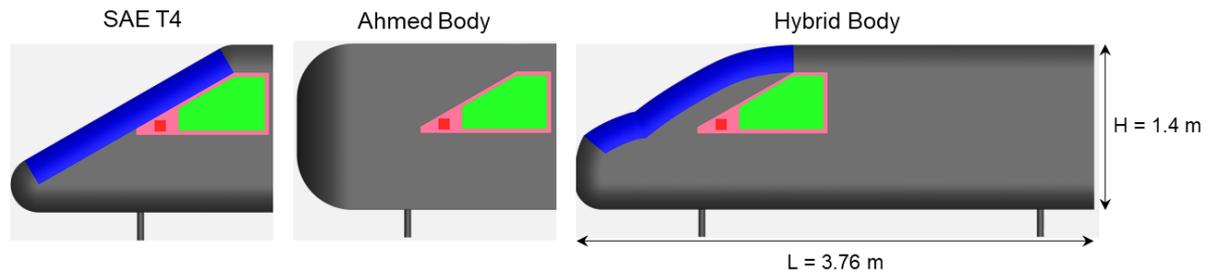

Figure.1: Schematic representation of all the models with used in this study with dimensions.

In the current study, a grid with wall normal units of ($\Delta x^+$ = 70 – 980; $\Delta y^+$ < 1, $\Delta z^+$ = 70 – 980) was developed in line with the previous study [10]. To predict the far-field noise, the Ffowcs Williams-Hawkings (FW-H) acoustic analogy was employed. All numerical simulations were carried out using ANSYS-Fluent (Version 2020) at Sheffield Hallam University's High-Performance Computing Cluster.

## 2   Results and discussion

A comparison of the hydrodynamic pressure fluctuations (HPF) predicted for the SAE-T4 case on the side window of various forebodies with experimental data from Nusser et al. (2021), at a location closest to the mirror position in Fig. 2a. Whilst the experimental data was only available for the SAE-Type 4 configuration from previous works [6,7], the SBES simulations of this configuration exhibited good agreement with the experimental data. Encouraged by these results, the simulation methodology and framework were extended to investigate the other two configurations. Therefore, it can be confidently asserted that the numerical simulations of all three configurations, although experimental data was only available for one, are reliable representations of the aerodynamic and aeroacoustics behavior of these configurations.

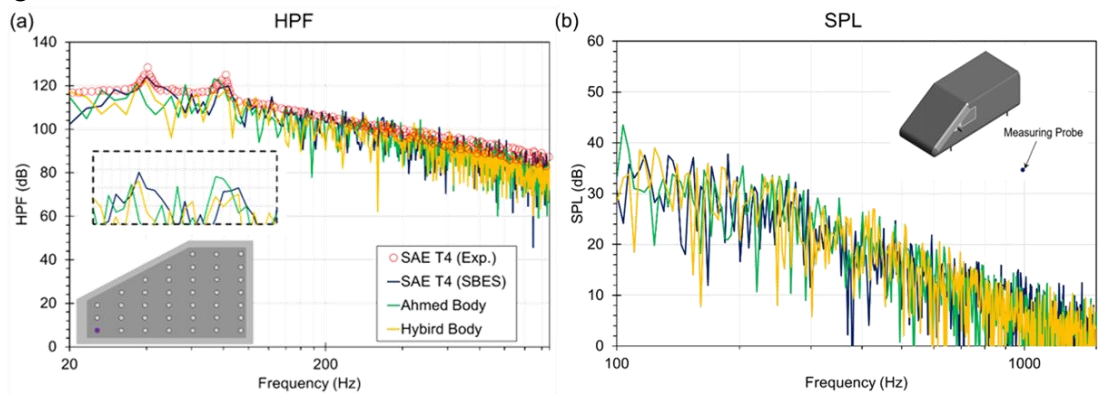

Figure.2: Comparison of (a) the predictions of the HPF on the side window at a location closest to the mirror from SBES for different forebody cases with experimental data of SAE-Type 4 from Nusser et al. (2021), (b) the predictions of the radiated sound numerical (SPL) from FW-H for different forebodies at a microphone located at 4m from the side window.

Moreover, Fig. 2a presents the results obtained for different forebody cases, indicating the presence of two distinct tonal peaks at approximately 40 Hz (peak-1) and 80 Hz (peak-2). These peaks correspond to Strohaul frequencies of $St \approx 0.116$ and $0.232$, respectively, where St is defined as $St = \frac{f.d}{U_\infty}$, with f representing the frequency of vortex shedding and *d* representing the characteristic length (side of the square) of the square cylinder.

Table 1: Comparison of force coefficients and noise radiated from all three geometries.

| **Geometry** | $C_d$ | $C_l$ | **OASPL** |
|---|---|---|---|
| SAE T4 | 0.2604 | -0.1266 | 60.46 dB |
| Ahmed Body | 0.2598 | -0.1774 | 60.12 dB |
| Hybrid Body | 0.2605 | -0.1326 | 59.93 dB |

While the peak-1 is more pronounced in the SAE-T4 case and the peak-2 is more prominent in the Ahmed body case, both distinct peaks are less pronounced in the Hybrid case compared to the other two cases. Notably, there are no distinct differences in the sound radiated from all three forebody cases, as shown in Figure 2b, and only negligible differences are observed in the predicted aerodynamic drag coefficient (*$C_d$*) and Overall (radiated) Sound Pressure Level (OASPL) at the microphone as shown in Table 1. However, significant differences in the predicted lift coefficient (*$C_l$*) are observed for all the three cases.

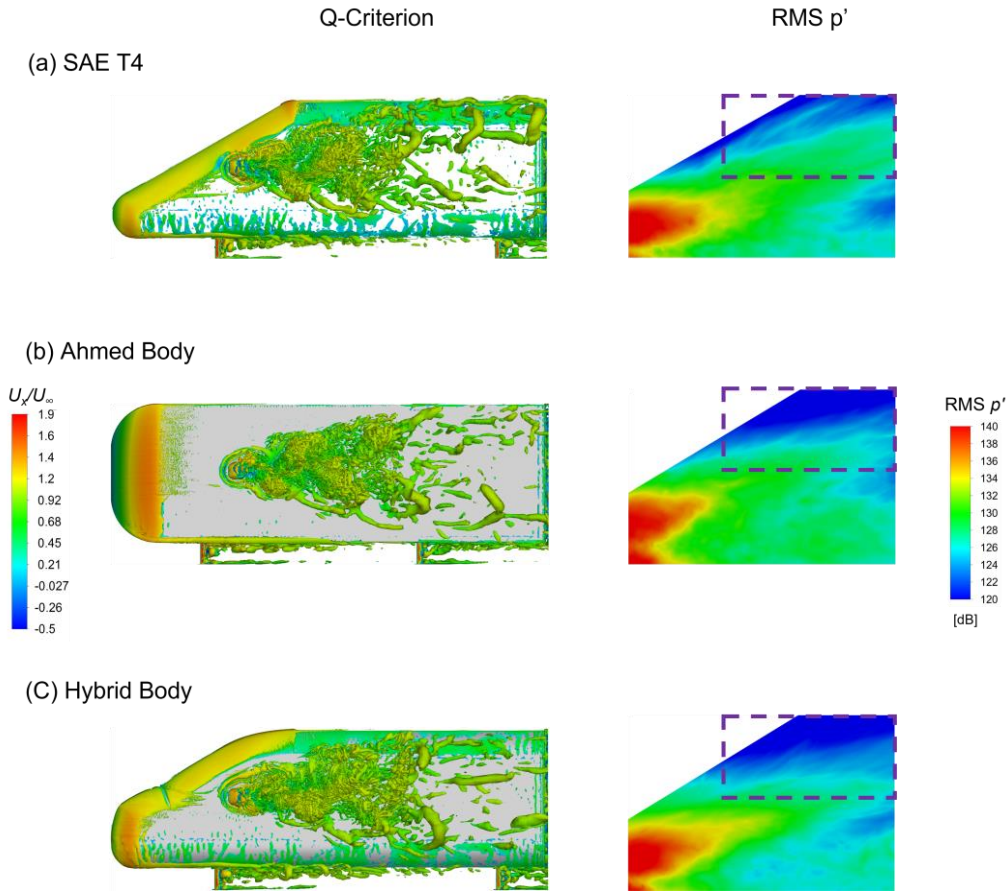

Figure.3: Comparison of vortical structures of an instantaneous flow field visualised by isosurfaces of Q = 1100 s$^{-2}$ coloured with instantaneous x-velocity (images on the left). On the right, the RMS of pressure fluctuations in dB on the side window for the three forebody cases.

To gain insight into the interaction of the A-pillar with the sideview mirror, the instantaneous flow structures for different forebody cases are compared, as shown in Figure 3 (images on the left). At the upstream of the mirror, the presence of a horseshoe vortex is evident, with highly unsteady smaller eddies generated downstream that interact with the A-pillar vortices, particularly in the case of SAE-T4 as compared to the Hybrid case, which inherits the features of both the Ahmed body and the SAE-T4. As a consequence, the overall distribution of pressure fluctuations on the side window, presented using the root-mean-square (RMS) of the pressure, is more pronounced for the SAE-T4 case as compared to the Hybrid case that has a smoother forebody curvature, as shown in Figure 3 (images on the right). Notably, this effect is not present in the Ahmed body, which does not have an A-pillar. In the Ahmed body case, the pressure fluctuations on the side window appear to be purely a contribution of the mirror, with an increased level of pressure fluctuations closer to the mirror and significantly reduced fluctuations closer to the upper portion of the side window than compared to the SAE-T4. This is the first study of its kind, and future work will examine additional forebody cases such as the Windsor body. It is envisaged that such a study shall provide a comprehensive understanding of the mechanism behind noise generation on the side window, as well as the noise radiated and aerodynamic forces from different forebody topologies.

## Acknowledgements

The Authors wish to thank Dr Kevin Chow of Horiba-MIRA, UK, and Dr Hauke Reese of ANSYS, Germany for useful discussions with the SAE Type 4 modelling.